\documentclass[twocolumn,showpacs,preprintnumbers,amsmath,amssymb]{revtex4}

\usepackage{graphicx}
\usepackage{dcolumn}
\usepackage{bm}
\usepackage{epsf}
\usepackage{psfrag}

\renewcommand{\bar}[1]{\overline{#1}}
\newcommand{\ie}{{\it i.e.}}

\usepackage{amssymb}

\usepackage{indentfirst}
\usepackage{psfig,color}
\usepackage{epsfig}
\usepackage{epsf}
\usepackage{graphicx}

\providecommand{\Journal}[4] {#1 {\bf #2}, #3 (#4)}
\providecommand{\EPJA}{Eur. Phys. J. A } %
\providecommand{\MPLA}{Mod. Phys. Lett. A} %
\providecommand{\NPA}{Nucl. Phys. A } %
\providecommand{\NPB}{Nucl. Phys. B } %
\providecommand{\PLB}{Phys. Lett. B } %
\providecommand{\PRL}{Phys. Rev. Lett. } %
\providecommand{\PRD}{Phys. Rev. D } %
\providecommand{\RMP}{Rev. Mod. Phys. } %
\providecommand{\RMP}{Rev. Mod. Phys. } %
\providecommand{\ZPA}{Z. Phys. A } %
\begin{document}

\title{The 35-plet Baryons from Chiral Soliton Models }

\author{Bin Wu}
\affiliation{Department of Physics, Peking University, Beijing
100871, China}
\author{Bo-Qiang Ma}
\altaffiliation{Corresponding author}\email{mabq@phy.pku.edu.cn}
\affiliation{ CCAST (World Laboratory), P.O.~Box 8730, Beijing 100080, China\\
Department of Physics, Peking University, Beijing 100871, China}

\begin{abstract}
We investigate the 35-plet baryons from the chiral soliton models.
We find that the coupling constant for the decay of 35-plet
baryons with spin 3/2 to the decuplet baryons is surprisingly
small, but that for the decay to 27-plet baryons with spin 3/2 is
larger. We give all the masses and widths of 35-plet baryons with
spin 5/2 and suggest candidates for all nonexotic members from the
available particle listings. We also focus on $\Delta_{5/2}$ and
$\Theta_2$, which are the lightest two baryons of 35-plet with
spin 5/2 and with simplest minimal pentaquark configurations.
Calculations show that $\Gamma_{\Delta_{5/2}}<$380 MeV, compared
with the results from $SU(2)$ Skyrme Model
($\Gamma_{\Delta_{5/2}}>$800 MeV), and $\Gamma_{\Theta_{2}}<100$
MeV if we assume that their widths are dominated by two-body decay
and that $\Theta^+$ has a width $\Gamma_{\Theta^+}<25$ MeV.
\end{abstract}


\vfill

\vfill

\maketitle
\par
\section{INTRODUCTION}

Chiral soliton models, based on chiral symmetry and large $N_c$
limit QCD, has played a crucial role in the
observations~\cite{LEPS,DIAN,CLAS,SAPH,HERMES,more} of an exotic
baryon with a narrow width and a positive strangeness number S=+1,
a possible candidate for $\Theta^+$ with minimal pentaquark
 configuration $\left|uudd\bar{s}\right>$~\cite{GM99}, which is the lightest
member of the antidecuplet, predicted from chiral soliton models
(Skyrme model and chiral quark-soliton
model)~\cite{Mano,Chem,Pra,penta1,Diak,Weig,Yan}. In chiral
soliton models, the generalization of $SU(2)$ collective
quantization to the $SU(3)$ case, taken into considerations the
chiral symmetry breaking terms, can indeed give the experimentally
agreeable mass splitting between the octet and the decuplet
baryons~\cite{Guad,Bolt}. The antidecuplet with spin 1/2, pointed
out by Manohar~\cite{Mano} and Chemtob~\cite{Chem}, is the third
baryon multiplet, besides the octet and the decuplet, which we
have been familiar with for a long time. By identifying $N(1710)$
as a known member, Diakonov, Petrov, and Polyakov~\cite{Diak}
calculated both the masses and the widths of the antidecuplet
baryons from chiral soliton models, among which the lightest
member $\Theta^+$ was predicted to have a mass 1530~MeV and a
narrow width $\Gamma_{\Theta^+}<30$~MeV~\cite{Jaff}, which agree
surprisingly with experimental results.
\par Based on the large $N_c$ analysis, the mass
difference between the antidecuplet and the octet in the $SU(3)$
chiral symmetry limit is of $O(1)$, which seems to invalidate the
collective quantization and the spurious prediction of $\Theta^+$
from chiral soliton models~\cite{cohen}. However, by introducing
``exoticness" ($X$), the minimal number of additional
quark-antiquark pairs needed to construct a multiplet on top of
the usual $N_c$ quarks, Diakonov and Petrov~\cite{diak2} showed
that the collective quantization description fails only when the
exoticness becomes comparable to $N_c$ and that in the case of the
antidecuplet the exoticness is 1, which is in favor of the
collective quantization. Moreover, under large $N_c$ limit, the
width of $\Theta^+$ is $1/N_c$ suppressed with respect to
$\Delta$~\cite{Pras}, which explains why $\Theta^+$ has so narrow
a width compared with other ordinary baryons from the point of
view of chiral soliton models.
\par In chiral soliton models, the allowed baryon multiplets are
those satisfying $I=J$ for hypercharge $Y=1$ baryons. Under such a
constraint, the baryon multiplets with exoticness=1, besides the
antidecuplet with spin 1/2, are 27-plets with spin 3/2 and 1/2 and
35-plets with spin 5/2 and 3/2, which, in the quark language, are
of the minimal five-quark configuration, \ie, so-called pentaquark
states. In Refs.~\cite{Wall,Bori,wu_ma1,wu_ma2,EKP,Pras27}, the
properties of those baryons were also discussed, and in
Ref.~\cite{wu_ma1} it has been found surprisingly that all the
nonexotic members of the 27-plet with spin 3/2 can be identified
from the available baryon listings~\cite{PDG}, agreeable with
experiments both in mass and width.
\par Because $J=\frac{5}{2}$ is
comparable with $N_c=3$, in the $SU(2)$ Skyrme model, the
higher-spin baryons ($I=J\geq5/2$) have so large angular
velocities that strong $\pi$ radiation will cause them to possess
a width $\Gamma>800$~MeV, dropping out of baryon mass
spectra~\cite{dhm}. However, in the $SU(3)$ case, the rotation is
distributed among more axes in the flavor space, which causes the
individual angular velocity to be smaller and, therefore, baryons
with spin 5/2 will be expected to have a narrower width, but a
higher mass. For large $N_c$, since mesons ($q\bar{q}$) and
baryons ($qqq$) are colorless, baryons are constituted of at least
$N_c$ quarks. For specific exoticness $X$, if we assume physical
baryon multiplets $(p,q)$ satisfying $I=J$ will maintain $X$ and
$J$ for arbitrarily large $N_c$, we can choose
\begin{eqnarray}
(p,q)=\left(p,\frac{N_c+3X-p}{2}\right),
\end{eqnarray}
with $p$ remains the same as the case of $N_c=3$, and~\cite{diak2}
\begin{eqnarray}
J_{\mbox{max}}=\frac{1}{6}(4p+2q-N_c).
\end{eqnarray}
Such a choice enable us to obtain the 35-plets with spin 5/2 and
3/2 $(4,\frac{N_c-1}{2})$ in the large $N_c$ limits, as well as
the octet $(1,\frac{N_c-1}{2})$, the decuplet
$(3,\frac{N_c-3}{2})$~\cite{largenc}, the antidecuplet
$(0,\frac{N_c+3}{2})$ and the 27-plets
$(2,\frac{N_c+1}{2})$~\cite{cohen,diak2}. And in the large $N_c$
limit, we have
\begin{eqnarray}
E_{35}-E_{\bar{10}}\longrightarrow 0.
\end{eqnarray}
Therefore the application of the chiral soliton models to the
35-plet with spin 5/2 is also valid if it is the case for the
antidecuplet as argued in Ref.~\cite{diak2}.
\par In this paper, we
calculate the masses and the widths of baryons in the 35-plets
from chiral soliton models, which are the last ones to be possibly
associated with pentaquark states in the quark language. We also
discuss $\Delta_{5/2}$ and $\Theta_2$ (or $\Theta^{**}$) with spin
5/2 in details, which are the lightest two members with simplest
minimal pentaquark configurations in the 35-plets. Our propose is
to give some experimentally testable results about the 35-plets
from chiral soliton models. This paper is organized as follows. In
Sec.~II, we briefly review the $SU(3)$ chiral soliton models. And
we give all the masses and the widths of the 35-plet baryons in
Sec.~III. In Sec.~IV, we calculate the widths of $\Delta_{5/2}$
and $\Theta_2$ up to linear order of $m_s$ and 1/$N_c$. And we
present our conclusion in Sec.~V. We list in Appendix the
Clebsch-Gordan coefficients involved with the products of the
$SU(3)$ irreducible representations 8(1,1) and 35(4,1).
\section{the SU(3) chiral soliton models}
In the $SU(3)$ chiral soliton models, the classical soliton, which
describes baryons, are of the form~\cite{Guad}
\begin{equation}
    U_1(\mathbf{x})=\left(
        \begin{array}{cc}
            \exp{[i(\mathbf{\widehat{r}}\cdot\mathbf{\tau})F(r)]} & \begin{array}{c}0\\0\end{array}\\
            \begin{array}{cc}0&0\end{array}&1
        \end{array}
        \right),
\end{equation}
where $F(r)$ is the spherical-symmetric profile of the soliton,
$\mathbf{\tau}$ are the three Pauli matrices, and
$\mathbf{\widehat{r}}$ is the unit vector in space. And
pseudoscalar fields can be written in collective coordinate as
\begin{equation}
    U(x)=\exp{\left[i\frac{\lambda_b\phi_b(x)}{f_\pi}\right]}=A(t)U_1(\mathbf{x})A(t)^{-1},\ \  A \in SU(3),
\end{equation}
and after quantizing on the collective coordinate $A$, we get the
$SU(3)$ chiral symmetry Hamiltonian~\cite{Guad}
\begin{equation}
\begin{array}{lr}
H_0=&M_{cl}+\frac{1}{2I_{1}}\sum\limits_{a=1}^3R_aR_a+\frac{1}{2I_{2}}\sum\limits_{b=4}^7R_bR_b,
\end{array}
\end{equation}
where $R_a$ are the angular momentums conjugate to the angular
velocities $\omega^a$, defined by
$A^\dagger\partial_0A=i\frac{1}{2}\omega^a\lambda^a$, $B$ is the
baryon number and $R_8=\frac{N_cB}{2\sqrt{3}}$. And the
corresponding mass spectra of baryon multiplets are
\begin{equation}
\begin{array}{lr}
E^{(p,q)}_J=&M_{cl}+\frac{1}{6I_2}\left[p^2+q^2+pq+3(p+q)-\frac{1}{4}(N_cB)^2\right]\\
&+\left(\frac{1}{2I_1}-\frac{1}{2I_2}\right)J(J+1)\label{H},
\end{array}
\end{equation}
where $(p,q)$ denotes an irreducible representation of the $SU(3)$
group, $M_{cl}$, $I_1$ and $I_2$ are given by integrating out the
classical soliton, treated model-independently and fixed by
experimental data, $M_{cl}$ is the classical soliton mass, $I_1$
and $I_2$ are moments of inertia.
\par The states of the system
will correspond to the baryon states, and wave function
$\Psi_{\nu\nu^\prime}^{(\mu)}$ of baryon $B$ in the collective
coordinates is of the form
\begin{equation}
    \Psi_{\nu\nu^\prime}^{(\mu)}(A)=\sqrt{\mbox{dim}(\mu)}D^{(\mu)}_{\nu\nu^\prime}(A),
\end{equation}
where $(\mu)$ denotes an irreducible representation of the SU(3)
group; $\nu$ and $\nu^{\prime}$ denote $(Y, I, I_3)$ and $(1, J,
-J_3)$ quantum numbers collectively; $Y$ is the hypercharge of
$B$; $I$ and $I_3$ are the isospin and its third component of $B$
respectively; $J_3$ is the third component of spin $J$;
$D^{(\mu)}_{\nu\nu^\prime}(A)$ are $SU(3)$ Wigner $D$-functions.
Due to the non-zero strange quark mass, the symmetry breaking
Hamiltonian is~\cite{war}
\begin{eqnarray}
    &&H^\prime=\alpha D^{(8)}_{88}+\beta
    Y+\frac{\gamma}{\sqrt{3}}\sum_{i=1}\limits^3D^{(8)}_{8i}J^i,\label{Hp}
\end{eqnarray}
where the coefficients $\alpha$, $\beta$, $\gamma$ are
proportional to the strange quark mass and model dependent;
$D^{(8)}_{ma}(A )$ is the adjoint representation of the SU(3)
group and defined as:
\begin{equation}
D^{(8)}_{ma}(A
)=\frac{1}{2}\mbox{Tr}(A^{\dagger}\lambda^mA\lambda^a).
\end{equation}
By introducing symmetry breaking term we can calculate the mass
splitting between baryons of the same baryon multiplets and the
physical wave functions. Yabu and Ando~\cite{ya} proposed a new
method to treat the symmetry breaking exactly. However, in this
paper, we treat the symmetry breaking term by perturbation theory,
and in the case of the octet and the decuplet, it can give
experimentally agreeable results~\cite{Bolt,Diak}.
\par In soliton
models, pseudoscalar Yukawa coupling for the process $B\rightarrow
B^\prime m$ can be obtained by Goldberger-Treiman relation, which
relates the relevant coupling constant to the axial
charge~\cite{Adki,Blot1}. And up to $1/N_c$ order, the coupling
operator in the space of the collective coordinates $A$ has the
form~\cite{Diak,Blot1}:
\begin{equation}
    \widehat{g}^{(m)}_A=G_0D^{(8)}_{m3}
    -G_1d_{3ab}D^{(8)}_{ma}J_b
    -\frac{G_2}{\sqrt{3}}D^{(8)}_{m8}J_3,\label{ga}
\end{equation}
where $d_{iab}$ is the SU(3) symmetric tensor, $a, b=4,5,6,7$, and
$J_a=-R_a$. $G_{1}$, $G_{2}$ are dimensionless constants, $1/N_c$
suppressed relative to $G_0$. And the width formula can be
obtained by~\cite{Diak,wu_ma2}
\begin{eqnarray}
    \Gamma(B\rightarrow B^\prime m)=\frac{3g^2_{BB^\prime
m}}{4\pi
m_B}|\mathbf{p}|\left[(m_{B^\prime}^2+\mathbf{p}^2)^{\frac{1}{2}}-m_{B^\prime}\right],
\end{eqnarray}
where $g^2_{BB^\prime m}$ can be calculated from
$g^{(m)}_A$~\cite{Diak,wu_ma2}.
\section{the 35-plet from chiral soliton models}
\begin{figure}
\vspace{0.3cm}
\begin{center}
\includegraphics[width=8cm]{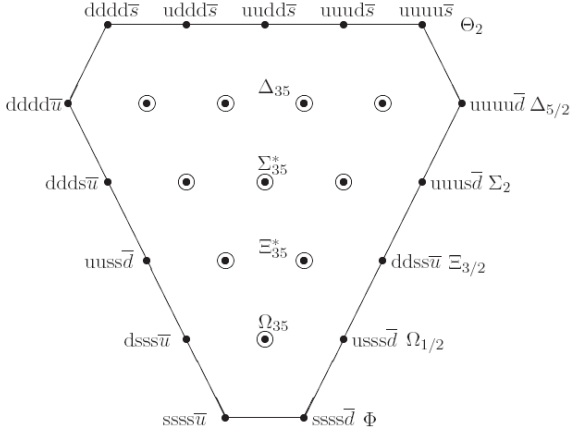}
\end{center}
\vspace{0.25cm} \caption{The quark content of the 35-plet
baryons.}\label{fig1}
\end{figure}

In the 35-plet, there are $I=5/2$ and $I=3/2$ baryons with $Y=1$,
from which we know there are two 35-plets with $J=5/2$ and $J=3/2$
respectively. We suggest the minimal quark configurations of the
35-plets in Fig.~\ref{fig1}, and the mass splittings are list in
Table~\ref{tab1}. To fix the parameters model-independently, we
choose the second group of data in Refs.~\cite{wu_ma1,wu_ma1},
which are consistent with the reported narrow $\Xi^-\pi^-$ baryon
resonance with mass of $1.862\pm0.003$~GeV and width below the
detector resolution of about $0.018$~GeV~\cite{NA49}
\begin{eqnarray}
\begin{array}{lll}\label{para}
\alpha=-663~\mbox{MeV},&
\beta=-12~\mbox{MeV},&\gamma=185~\mbox{MeV},\\
1/I_1=154~\mbox{MeV},&1/I_2=399~\mbox{MeV},&M_{cl}=798~\mbox{MeV}.
\end{array}\end{eqnarray}

\begin{widetext}
\begin{center}
\begin{table}
\caption{\label{tab1}The masses (GeV) of baryons in the 35-plets}
\begin{ruledtabular}
\begin{tabular}{cccccccc}
& $Y$ & $I$ & $\left<B|H^\prime|B\right>$&Predicted Mass&Candidate&$I(J^{P})$&PDG\\
\hline
35-plet $J=5/2$\footnotemark[1]\\
\hline
$\Delta$ & 1 &$\frac{3}{2}$ &$-\frac{1}{8}\alpha+\beta -\frac{7}{16}\gamma$&1.96&$\Delta(1905)$&$\frac{3}{2}(\frac{5}{2}^+)$&$1.87~\mbox{to}~1.92$\\
$\Sigma$      &0& 1   &$-\frac{3}{16}\alpha-\frac{7}{96}\gamma$&2.08&$\Sigma(2070)$&$1(\frac{5}{2}^+)$&$\approx2.07$\\
$\Xi$     &-1& $\frac{1}{2}$ &$-\frac{1}{4}\alpha -\beta +\frac{7}{24}\gamma$&2.20&$\Xi(2250)$&$\frac{1}{2}(?^?)$&$\approx2.25$\\
$\Omega$      &-2& 0   &$-\frac{5}{16}\alpha -2\beta +\frac{21}{32}\gamma$&2.32&$\Omega(2250)^-$&$0(?^?)$&$2.252\pm0.009$\\
\hline
$\Theta_2$  &    2 & 2   &$-\frac{1}{16}\alpha +2\beta-\frac{77}{96}\gamma)$&1.84&?&$?(?^?)$&?\\
$\Delta_{5/2}$ & 1 & $\frac{5}{2}$&$\frac{11}{32}\alpha +\beta
-\frac{49}{192}\gamma$&1.68&?&$?(?^?)$&?\\£©
$\Sigma_2$&0 & 2   &$\frac{3}{16}\alpha +\frac{7}{96}\gamma$&1.86&?&$?(?^?)$&?\\
$\Xi_{3/2}$ &-1&$\frac{3}{2}$ &$\frac{1}{32}\alpha -\beta +\frac{77}{192}\gamma$&2.04&?&$?(?^?)$&?\\
$\Omega_1$      &-2& 1   &$-\frac{1}{8}\alpha-2\beta +\frac{35}{48}\gamma$&2.21&?&$?(?^?)$&?\\
$\Phi$          &-3& $\frac{1}{2}$ &$-\frac{9}{32}\alpha -3\beta +\frac{203}{192}\gamma$&2.39&?&$?(?^?)$&?\\
\hline
35-plet $J=3/2$\\
\hline
$\Delta$ & 1 &$\frac{3}{2}$ &$\frac{3}{16}\alpha+\beta +\frac{1}{32}\gamma$&2.45&?&$?(?^?)$&?\\
$\Sigma$      &0& 1   &$\frac{1}{8}\alpha+\frac{3}{16}\gamma$&2.53&?&$?(?^?)$&?\\
$\Xi$     &-1& $\frac{1}{2}$ &$\frac{1}{16}\alpha -\beta +\frac{11}{32}\gamma$&2.62&?&$?(?^?)$&?\\
$\Omega$      &-2& 0   &$ -2\beta +\frac{1}{2}\gamma$&2.70&?&$?(?^?)$&?\\
\hline
$\Theta_2$  &    2 & 2   &$\frac{1}{4}\alpha +2\beta-\frac{1}{8}\gamma)$&2.37&?&$?(?^?)$&?\\
$\Delta_{5/2}$ & 1 & $\frac{5}{2}$&$-\frac{1}{8}\alpha +\beta -\frac{7}{16}\gamma$&2.57&?&$?(?^?)$&?\\
$\Sigma_2$&0 & 2   &$\frac{3}{16}\alpha +\frac{7}{96}\gamma$&2.63&?&$?(?^?)$&?\\
$\Xi_{3/2}$ &-1&$\frac{3}{2}$ &$-\frac{1}{8}\alpha -\beta +\frac{1}{16}\gamma$&2.69&?&$?(?^?)$&?\\
$\Omega_1$      &-2& 1   &$-\frac{1}{8}\alpha-2\beta +\frac{5}{16}\gamma$&2.75&?&$?(?^?)$&?\\
$\Phi$          &-3& $\frac{1}{2}$ &$-\frac{1}{8}\alpha -3\beta +\frac{9}{16}\gamma$&2.81&?&$?(?^?)$&?\\
\end{tabular}
\end{ruledtabular}
\footnotetext[1]{already calculated in Ref.~\onlinecite{Bori}.}
\end{table}
\end{center}
\end{widetext}

\par Due to the symmetry breaking Hamiltonian (\ref{Hp}), the
physical baryon wave functions will mix with those belonging to
other baryon multiplets but with the same spin and isospin. For
the decuplet and the 35-plet baryons, by first perturbation, we
have
\begin{eqnarray}
        \left|B^{(10)}\right>&=&\left|B;10\right>+a_{27}\left|B;27_{\frac{3}{2}}\right>+a_{35}\left|B;35_{\frac{3}{2}}\right>\\\nonumber
        \left|B^{(35)}_{\frac{5}{2}}\right>&=&\left|B;35_{\frac{5}{2}}\right>+b_{28}\left|B;28\right>\\
        &+&b_{64}\left|B;64_\frac{5}{2}\right>+b_{81}\left|B;81_\frac{5}{2}\right>,\\\nonumber
         \left|B^{(35)}_{\frac{3}{2}}\right>&=&\left|B;35_{\frac{3}{2}}\right>+c_{10}
        \left|B;10\right>+c_{27}\left|B;27_\frac{3}{2}\right>\\
        &+&c_{64}\left|B;64_\frac{3}{2}\right>+c_{81}\left|B;81_\frac{3}{2}\right>,
\end{eqnarray}
and the coefficients above are given simply by perturbation
theory. To calculate the widths of the 35-plet baryons, we first
calculate the coupling constant $g^2_{BB^\prime m}$ from
$\widehat{g}^{(m)}_A$. From (\ref{ga}), we have
\begin{eqnarray}\nonumber
    \widehat{g}^{(m)}_A&=&G_0D^{(8)}_{m3}
    +\frac{\sqrt{2}}{4}G_1\left(D_{mn}^{(8)}R_{6+7i}+D_{m\Xi^0}^{(8)}R_{6-7i}\right.\\
    &-&\left.D_{mp}^{(8)}R_{4+5i}+D_{m\Xi^-}^{(8)}R_{4-5i}\right)-\frac{G_2}{\sqrt{3}}D^{(8)}_{m8}J_3,\label{ga1}
\end{eqnarray}
where $D_{mB^{(8)}}^{(8)}=\left<B^{(8)}|D^{(8)}|m\right>$ and
$R_{a\pm bi}\equiv R_a\pm iR_b$. Trivial algebra will give us
\begin{eqnarray}
\left<B;35_{\frac{5}{2}}|\widehat{g}^{(m)}_A|B^\prime;10\right>&=&G_{35_{\frac{5}{2}}10}\left<B;35_\frac{5}{2}|D^{(8)}_{m3}|B^\prime;10\right>,\\
\left<B;35_{\frac{5}{2}}|\widehat{g}^{(m)}_A|B^\prime;27\right>&=&G_{35_{\frac{5}{2}}27}\left<B;35_\frac{5}{2}|D^{(8)}_{m3}|B^\prime;27\right>,\\
\left<B;35_{\frac{3}{2}}|\widehat{g}^{(m)}_A|B^\prime;10\right>&=&G_{35_{\frac{3}{2}}10}\left<B;35_\frac{3}{2}|D^{(8)}_{m3}|B^\prime;10\right>,\\
\left<B;35_{\frac{3}{2}}|\widehat{g}^{(m)}_A|B^\prime;27\right>&=&G_{35_{\frac{3}{2}}27}\left<B;35_\frac{3}{2}|D^{(8)}_{m3}|B^\prime;27\right>,
\end{eqnarray}
with
\begin{eqnarray}\nonumber
G_{35_{\frac{5}{2}}10}&=&G_0,G_{35_{\frac{5}{2}}27}=G_0+G_1\\\nonumber
G_{35_{\frac{3}{2}}10}&=&G_0-\frac{5}{2}G_1+\frac{5}{2}G_2,\\\nonumber
G_{35_{\frac{3}{2}}27}&=&G_0-\frac{1}{14}G_1+\frac{15}{14}G_2.
\end{eqnarray}
Compared with the coupling constant for the decay of antidecuplet
baryons $G_{\bar{10}}=G_0-G_1-1/2G_2$ and the possible narrow
width of $\Theta^+$, the width of the 35-plet baryons with spin
3/2 will be surprisingly suppressed since in chiral quark-soliton
model the ration $G_1$/$G_0$ ranges from 0.4 to 0.6~\cite{Blot1}.
Similarly for the decay of 27-plet baryons with spin 1/2 to the
octet, the coupling is also small
$G_{27}=G_0-2G_1+\frac{3}{2}G_2$~\cite{Pras27}. And in this paper,
we will focus only on the 35-plet with spin 5/2. Under the chiral
$SU(3)$ symmetry, we have the following width
formula~\cite{wu_ma1}
\begin{widetext}\begin{center}
\begin{equation}
    \Gamma(B\rightarrow B^\prime m)=\frac{G_s^2}{4\pi}\frac{|\mathbf{p}|}{m_B}
    \left[(m_{B^\prime}^2+\mathbf{p}^2)^{\frac{1}{2}}-m_{B^\prime}\right]\left\{\begin{array}{c}
    \frac{\mbox{dim}(\mu^\prime)}{\mbox{dim}(\mu)}\left|\begin{array}{c}\sum\limits_\gamma\left(
    \begin{array}{cc}8&\mu^\prime\\ Y_mI_m&Y_\rho I_\rho \end{array}\right|
    \left.\begin{array}{c}\mu_\gamma\\Y_\nu
    I_\nu\end{array}\right)
    \left(
    \begin{array}{cc}8&\mu^\prime\\01&1 J_\rho \end{array}\right|
    \left.\begin{array}{c}\mu_\gamma\\1J_\nu\end{array}\right)\end{array}\right|^2\end{array}\right\},
\end{equation}
\end{center}\end{widetext}
where we postulate $B$ with $(Y, I, I_3; J^P, -J_3)=(Y_\nu, I_\nu,
I_{\nu3}; J_\nu^+, -J_{\nu3})$, $B^\prime$ with $(Y, I, I_3; J^P,
-J_3)=(Y_\rho, I_\rho, I_{\rho3}; J_\rho^+, -J_{\rho3})$ and $m$
with $(Y, I, I_3; J^P, -J_3)=(Y_m, I_m, I_{m3}; 0^-, 0)$; and
$G_s^2=3.84G_{35}^2$. Using this formula, and provided that the
width of $\Theta^+~$ $\Gamma_{\Theta^+}<25~\mbox{MeV}$, we
calculate the upper bounds of widths for decay of the 35-plet
baryons with spin 5/2 to the decuplet baryons, listed in
Table~\ref{tab2}. From the masses, $I(J^P)$ and widths, we suggest
candidates for all nonexotic members of the 35-plet with spin 5/2
from the available particle listings~\cite{PDG}. For $\Delta$,
$\Sigma$ and $\Xi$, it is interesting to see that our calculation,
though not as good as those 27-plet nonexotic
candidates~\cite{wu_ma1}, are still consistent with experimental
results, but the $\Omega$ candidate is not so ideal. Therefore, we
may suggest that $\Xi(2250)$, as a member of the 35-plet, is with
the quantum numbers $I(J^P)$=$\frac{1}{2}(\frac{5}{2}^+)$.
\begin{widetext}
\begin{center}
\begin{table}
\caption{\label{tab2}The widths (MeV) of baryons in the 35-plet
with spin 5/2}
\begin{ruledtabular}
\begin{tabular}{ccccc}
\multicolumn{5}{c}{}\\\hline 35-plet baryons& modes&width
$\leq$calculation&total width&PDG data
\\\hline
$\Delta$&$\Delta\pi$&42&&\\
&$\Delta\eta$&63&135&$280~\mbox{to}~400$\\
&$\Sigma^* K$&30&&\\\hline
$\Sigma$&$\Sigma^*\pi$&44&&\\
&$\Sigma^*\eta$&65&161&$(300\pm 30)/906/(140\pm 20)$\footnotemark[2]\\
&$\Xi^*K$&9&&\\
&$\Delta\bar{K}$&43&&\\\hline
$\Xi$&$\Xi^*\pi$&27&&\\
&$\Xi^*\eta$&49&112&$(46\pm 27)/(<30)/(130\pm 80)$\footnotemark[2]\\
&$\Omega K$&1&&\\
&$\Sigma^*\bar{K}$&35&&\\\hline
$\Omega$&$\Omega\pi$&29&175&$55\pm 18$\\
&$\Xi^*\bar{K}$&146&&\\\hline

$\Theta_2$&$\Delta K$&93&93&?\\\hline
$\Delta_{5/2}$&$\Sigma\pi$&206&206&?\\\hline
$\Sigma_2$&$\Sigma^*\pi$&154&154&?\\
&$\Delta\bar{K}$&30&30&\\\hline
$\Xi_{3/2}$&$\Xi^*\pi$&108&176&?\\
&$\Sigma^*\bar{K}$&68&&\\\hline
$\Omega_1$&$\Omega\bar{K}$&180&180&?
\end{tabular}
\end{ruledtabular}
\footnotetext[2]{Data from different Collaborations, and PDG
provided no estimations.}
\end{table}
\end{center}
\end{widetext}

\section{discussion of $\Delta_{5/2}$ and $\Theta_2$}
From Table \ref{tab1}, we can see that $\Delta_{5/2}$ are the
lightest baryons, and among them, there are $\Delta_{5/2}^{--}$
and $\Delta_{5/2}^{+++}$, which, in the quark language, are with
the simplest minimal pentaquark configurations
$\left|dddd\bar{u}\right>$ and $\left|uuuu\bar{d}\right>$
respectively~\cite{GM99}. And in the 35-plets, $\Theta_2$ states
with isospin 2, the excitations of $\Theta^+$, include those
baryons with minimal pentaquark configurations
$\left|dddd\bar{s}\right>$ and $\left|uuuu\bar{s}\right>$. In
Sec.~III, we only give the width of 35-plet baryons by the
simplification of chiral $SU(3)$ symmetry, and we will calculate
the widths of $\Delta_{5/2}$ and $\Theta_2$ up to linear order of
$m_s$ and 1/$N_c$ below.
\par From (\ref{Hp}), the physical baryon states are of
the form by first-order approximation
\begin{widetext}
\begin{eqnarray}\nonumber
        &\left|\Delta\right>&=\left|\Delta;10\right>+C_{27}^{(\Delta)}
        \left|\Delta;27\right>+C_{35}^{(\Delta)}\left|\Delta;35_\frac{3}{2}\right>,\\\nonumber
        &\left|\Delta^*\right>&=\left|\Delta^*;27_{\frac{3}{2}}\right>+C_{10}^{\Delta^*}\left|\Delta;10\right>+C_{\bar{35}}^{\Delta^*}
        \left|\Delta;\bar{35}_\frac{3}{2}\right>+C_{35}^{\Delta^*}
        \left|\Delta;35_\frac{3}{2}\right>+C_{64}^{\Delta^*}\left|\Delta;64_\frac{3}{2}\right>,\\\nonumber
        &\left|\Theta^{*}\right>&=\left|\Theta^{*};27_\frac{3}{2}\right>
        +C_{\bar{35}}^{(\Theta^{*})}\left|\Theta^{*};\bar{35}_\frac{3}{2}^{(\Theta^{*})}\right>+C_{64}^{(\Theta^{*})}\left|\Theta^{*};64_\frac{3}{2}\right>,
        \left|\Theta_2\right>=\left|\Theta_2;35_\frac{5}{2}\right>
        +C_{64}^{(\Theta_2)}\left|\Theta_2;64_\frac{5}{2}\right>+C_{81}^{(\Theta_2)}\left|\Theta_2;81_\frac{5}{2}\right>,\\\nonumber
        &\left|\Delta_{5/2}\right>&=\left|\Delta_{5/2};35_\frac{5}{2}\right>+C_{28}^{(\Delta_{5/2})}
        \left|\Delta_{5/2};28\right>+C_{64}^{(\Delta_{5/2})}\left|\Delta_{5/2};64_\frac{5}{2}\right>+C_{81}^{(\Delta_{5/2})}\left|\Delta_{5/2};81_\frac{5}{2}\right>,
\end{eqnarray}
with
\begin{eqnarray}\nonumber
C_{27}^{(\Delta)}&=&-\frac{\sqrt{30}}{16}\left(\alpha+\frac{5}{6}\gamma\right)I_2,
C_{35}^{(\Delta)}=-\frac{5\sqrt{14}}{336}\left(\alpha-\frac{1}{2}\gamma\right)I_2,
C_{10}^{(\Delta^*)}=\frac{\sqrt{30}}{16}\left(\alpha+\frac{5}{6}\gamma\right)I_2,\\\nonumber
C_{\bar{35}}^{(\Delta^*)}&=&-\frac{\sqrt{105}}{70}\left(\alpha+\frac{5}{6}\gamma\right)I_2,
C_{35}^{(\Delta^*)}=-\frac{\sqrt{105}}{224}\left(\alpha-\frac{7}{6}\gamma\right)I_2,
C_{64}^{(\Delta^*)}=-\frac{5\sqrt{3}}{196}\left(\alpha-\frac{1}{6}\gamma\right)I_2,\\\nonumber
C_{\overline{35}}^{(\Theta^{*})}&=&-\frac{3\sqrt{35}}{140}\left(\alpha+\frac{5}{6}\gamma\right)I_2,
C_{64}^{(\Theta^{*})}=-\frac{3\sqrt{10}}{196}\left(\alpha-\frac{1}{6}\gamma\right)I_2,
C_{64}^{(\Theta_2)}=-\frac{\sqrt{35}}{30}\left(\alpha+\frac{7}{6}\gamma\right)I_2,\\\nonumber
C_{81}^{(\Theta_2)}&=&-\frac{7\sqrt{15}}{960}\left(\alpha-\frac{1}{2}\gamma\right)I_2,
C_{28}^{(\Delta_{5/2})}=-\frac{\sqrt{5}}{48}\left(\alpha-\frac{7}{6}\gamma\right)I_2,
C_{64}^{(\Delta_{5/2})}=-\frac{\sqrt{35}}{30}\left(\alpha+\frac{7}{6}\gamma\right)I_2,\\\nonumber
C_{81}^{(\Delta_{5/2})}&=&-\frac{7\sqrt{35}}{1920}\left(\alpha-\frac{1}{2}\gamma\right)I_2.
\end{eqnarray}
Substituting the parameters (\ref{para}) into these coefficients,
we have
\begin{eqnarray}\nonumber
&C_{27}&=0.44,C_{35}=0.09,C_{10}^{(\Delta^*)}=-0.44,C_{\bar{35}}^{(\Delta^*)}=0.19,
C_{35}^{(\Delta^*)}=0.10, C_{64}^{(\Delta^*)}=0.08,\\\nonumber
&C_{\bar{35}}^{(\Theta^{*})}&=0.16,C_{64}^{(\Theta^{*})}=0.08,C_{64}^{(\Theta_2)}=0.22,C_{81}^{(\Theta_2)}=0.05,
C_{28}^{(\Delta_{5/2})}=0.10,C_{64}^{(\Delta_{5/2})}=0.22,C_{81}^{(\Delta_{5/2})}=0.04.
\end{eqnarray}
By the general width formula~\cite{Diak,wu_ma2}, for the decay of
$\Delta_{5/2}$ and $\Theta_2$ to the octet baryons, we have
\begin{eqnarray}
    \Gamma(\Theta_{2} \rightarrow \Delta K)&=&g_0\frac{G_0^2}{14\pi
m_{\Theta_{2}}}|\mathbf{p}|\left[(m_{\Delta}^2+\mathbf{p}^2)^{\frac{1}{2}}-m_{\Delta}\right]\left(1-\frac{\sqrt{30}}{5}C_{27}^{(\Delta)}\right)<65
~\mbox{MeV},~~~~~~\\\nonumber
    \Gamma(\Delta_{5/2} \rightarrow \Delta\pi)&=&g_0\frac{G_0^2}{14\pi
m_{\Delta_{5/2}}}|\mathbf{p}|\left[(m_{\Delta}^2+\mathbf{p}^2)^{\frac{1}{2}}-m_{\Delta}\right]\left(1+\frac{\sqrt{30}}{5}C_{27}^{(\Delta)}+\frac{\sqrt{14}}{10}C_{35}^{(\Delta)}\right)<380
~\mbox{MeV},~~~~~~\\\nonumber
\end{eqnarray}
and, also, to the 27-plet baryon $\Theta^*$~\cite{wu_ma1,wu_ma2}
\begin{eqnarray}
    \Gamma(\Theta_{2} \rightarrow \Theta^*\pi)&=&g_0\frac{3(G_0+G_1)^2}{70\pi
m_{\Theta_{2}}}|\mathbf{p}|\left[(m_{\Delta}^2+\mathbf{p}^2)^{\frac{1}{2}}-m_{\Delta}\right]\left(1+\frac{\sqrt{10}}{20}\frac{G_0}{G_0+G1}C_{64}^{(\Theta^*)}+\frac{\sqrt{35}}{4}\frac{G_0}{G_0+G1}C_{64}^{(\Theta_2)}\right)<35~\mbox{MeV},~~~~~~\\\nonumber
\end{eqnarray}
and other processes are prohibited by the conservations of energy
and momentum and the numeric values are estimated provided that
$\Gamma_{\Theta^+}<25~\mbox{MeV}$.
\end{widetext}
\section{summary and conclusions}
The baryon 35-plets with spin 5/2 and 3/2 are the last two $SU(3)$
multiples with exoticness $X=1$, which may be associated with
so-called pentaquark states. And we give all the masses of the
35-plet baryons (Table \ref{tab1}), and to estimate their widths,
we also calculate the coupling constants by Goldberger-Treiman
relation. We find that 35-plet with spin 3/2 have a even more
narrow coupling constant for the decay of those baryons to the
decuplet baryons than the antidecuplet, while that for the process
involved with the 27-plet baryons with 3/2 are relatively large.
In contrast, the coupling constants for the 35-plet with spin 5/2
are larger than that of the antidecuplet, which enable those
baryons to possess a much broader width than that of the
antidecuplet baryons, as listed in Table~\ref{tab2}.
\par To give experimentally testable results further, we calculate the
widths of $\Delta_{5/2}$ and $\Theta_2$ up to linear order of
$m_s$ and 1/$N_c$. $\Delta_{5/2}$, in the quark language, may be
the simplest pentaquark states of pentaquark configuration only
involving $u$ and $d$ quarks~\cite{GM99} and there are
$\Delta_{5/2}^{+++}(uuuu\bar{d})$ and
$\Delta_{5/2}^{--}(dddd\bar{u})$ in these isospin multiplet. And
due to the conservation of energy and momentum, the decay of
$\Delta_{5/2}$ to the 27-plet baryons are prohibited. Therefore,
those baryons can only be revealed by the decay processes
involving the decuplet baryons
\begin{eqnarray}
\Delta_{5/2}^{+++}\longrightarrow\Delta^{++}\pi^+, \nonumber\\
\Delta_{5/2}^{--}\longrightarrow\Delta^-\pi^-. \nonumber
\end{eqnarray}
According to the calculation in Sec.~IV, we estimate
$\Gamma_{\Delta_{5/2}}<$380~MeV if we choose
$\Gamma_{\Theta^+}<25$~MeV and assume that the width are dominated
by two-body decay. The $\Theta_2$ states, including
$\Theta_2^{+++}(uuuu\bar{s})$, $\Theta_2^{++}(uuud\bar{s})$,
$\Theta_2^{+}(uudd\bar{s})$, $\Theta_2^{0}(uddd\bar{s})$, and
$\Theta_2^{-}(dddd\bar{s})$, can decay to the decuplet baryons and
the 27-plet baryons with spin 3/2, \ie,
\begin{eqnarray}
\Theta_{2}\rightarrow \Delta K, \nonumber\\
\Theta_{2}\rightarrow \Theta^*\pi.
\end{eqnarray}
and the total
width $\Gamma_{\Theta_{2}}<100$~MeV on the assumption of
$\Gamma_{\Theta^+}<25$~MeV and dominance of two-body decay.

In summary, as argued by Diakonov,and Petrov~\cite{diak2}, chiral
soliton models are valid in the predictions about the exotic
baryon mulitplets when the exoticness is small comparable with
$N_c$. And among the baryon multiplets with exoticness=1, such as
the antidecuplet, 35-plets with spin 5/2 and 3/2 are the last
(most weighted) two baryon multiplets from the chiral soliton
models which may be described by pentaquark states in quark
language. And in this paper we investigate the 35-plet baryons and
suggest all nonexotic members with spin 5/2 from the available
particle listings. We focus on the baryons with spin 5/2,
especially $\Delta_{5/2}$ and $\Theta_2$, which are of simplest
pentaquark configurations and find that $\Delta_{5/2}$ have a
width much less than the prediction from the $SU(2)$ Skyrme Model,
which is physically testable. These two baryons both have a
typical width of strong decay, and the search of them are also of
great significance to test the validity of application of chiral
soliton models to these so-called exotic baryon multiplets in the
eyes of the quark model.

\section*{ACKNOWLEGEMENT}
This work is partially supported by
National Natural Science Foundation of China under Grant Numbers
10025523 and 90103007.
\section*{APPENDIX: The Clebsch-Gordan Coefficients for the product of $35(4,2)\otimes8(1,1)$}
In this paper, we use frequently the Clebsch-Gordan Coefficients
for the product of the $SU(3)$ irreducible representations
$35(4,2)\otimes8(1,1)$, which decomposes as

\begin{eqnarray}\nonumber
&35&(4,2)\otimes8(1,1)=10(3,0)\oplus 28(6,0)\oplus 27(2,2)\\
&\oplus& 35^{(1)}(4,1)\oplus 35^{(2)}(4,1)\oplus 64(3,3)\oplus
81(5,2).
\end{eqnarray}

To be convenient, we define
\begin{eqnarray}
\mu^{(\lambda)}_{(\nu)}=\sum\limits_{\nu_1,\nu_2}
\left(\begin{array}{ccc}\mu_1&\mu_2&\mu\\\nu_1&\nu_2&\nu\end{array}\right)\mu_{1(\nu_1)}\mu_{2(\nu_2)},
\end{eqnarray}
where $\mu_{(\nu)}^{\lambda}$ denote the eigenstates of the
representation $\mu$ contained in the direct sum of $\mu_1$ and
$\mu_2$, whose eigenstates are $\mu_{1(\nu_1)}$ and
$\mu_{2(\nu_2)}$ respectively, $\lambda$ is used to distinguish
identical but independent representations which are all contained
in $\mu_1\otimes\mu_2$, $\nu$,$\nu_1$ and $\nu_2$ denote quantum
number $(YII_3)$ collectively, and
$\left(\begin{array}{ccc}\mu_1&\mu_2&\mu\\\nu_1&\nu_2&\nu\end{array}\right)$
are the $SU(3)$ Clebsch-Gordan coefficients. We define~\cite{Swar}
\begin{eqnarray}
I_{\pm}=F_1\pm iF_2,\label{Ipm}\\
K_{\pm}=F_4\pm iF_5,\\
L_{\pm}=F_6\pm iF_7,\label{Lpm}
\end{eqnarray}
where $F_a$ are the generators of the $SU(3)$ group. Following the
conventions used in Ref. \cite{Swar}, we choose all the
coefficients of any states given by the action of $I_{\pm}$ and
$K_{\pm}$ on any eigenstates of $F_3$ and $F_8$ are nonnegative
and give the follow decomposition
\begin{widetext}
\begin{eqnarray}\nonumber
&81_{3\frac{5}{2}\frac{5}{2}}&=35_{222}8_{1\frac{1}{2}\frac{1}{2}},\\\nonumber
&81_{222}&=\frac{\sqrt{30}}{30}35_{222}8_{010}+\frac{3\sqrt{10}}{20}35_{222}8_{000}
-\frac{\sqrt{15}}{30}35_{221}8_{011}-\frac{\sqrt{15}}{60}35_{1\frac{5}{2}\frac{5}{2}}8_{1\frac{1}{2}-\frac{1}{2}}
+\frac{\sqrt{3}}{60}35_{1\frac{5}{2}\frac{3}{2}}8_{1\frac{1}{2}\frac{1}{2}}
+\frac{3\sqrt{2}}{5}35_{1\frac{3}{2}\frac{3}{2}}8_{1\frac{1}{2}\frac{1}{2}},\\\nonumber
&81_{1\frac{5}{2}\frac{5}{2}}&=\frac{\sqrt{35}}{20}35_{222}8_{-1\frac{1}{2}\frac{1}{2}}
+\frac{3\sqrt{70}}{280}35_{1\frac{5}{2}\frac{5}{2}}8_{010}
+\frac{\sqrt{210}}{40}35_{1\frac{5}{2}\frac{5}{2}}8_{000}
-\frac{3\sqrt{7}}{140}35_{1\frac{5}{2}\frac{3}{2}}8_{011}+\frac{\sqrt{42}}{10}35_{1\frac{3}{2}\frac{3}{2}}8_{011}
+\frac{\sqrt{35}}{10}35_{022}8_{1\frac{1}{2}\frac{1}{2}},\\\nonumber
&64_{222}&=\frac{\sqrt{30}}{15}35_{222}8_{010}-\frac{\sqrt{10}}{5}35_{222}8_{000}
-\frac{\sqrt{15}}{15}35_{221}8_{011}+\frac{2\sqrt{15}}{15}35_{1\frac{5}{2}\frac{5}{2}}8_{1\frac{1}{2}-\frac{1}{2}}
-\frac{2\sqrt{3}}{15}35_{1\frac{5}{2}\frac{3}{2}}8_{1\frac{1}{2}\frac{1}{2}}
+\frac{\sqrt{2}}{5}35_{1\frac{3}{2}\frac{3}{2}}8_{1\frac{1}{2}\frac{1}{2}},\\\nonumber
&64_{211}&=\frac{\sqrt{105}}{35}35_{222}8_{01-1}-\frac{\sqrt{210}}{70}35_{221}8_{010}
+\frac{\sqrt{70}}{70}35_{221}8_{011}
+\frac{3\sqrt{14}}{14}35_{1\frac{3}{2}\frac{3}{2}}8_{1\frac{1}{2}-\frac{1}{2}}
-\frac{\sqrt{42}}{14}35_{1\frac{3}{2}\frac{1}{2}}8_{1\frac{1}{2}\frac{1}{2}},\\\nonumber
&64_{1\frac{5}{2}\frac{5}{2}}&=-\frac{2\sqrt{15}}{15}35_{222}8_{-1\frac{1}{2}\frac{1}{2}}
+\frac{\sqrt{30}}{15}35_{1\frac{5}{2}\frac{5}{2}}8_{010}
-\frac{\sqrt{10}}{5}35_{1\frac{5}{2}\frac{5}{2}}8_{000}
-\frac{2\sqrt{3}}{15}35_{1\frac{5}{2}\frac{3}{2}}8_{011}+\frac{\sqrt{2}}{5}35_{1\frac{3}{2}\frac{3}{2}}8_{011}
+\frac{\sqrt{15}}{15}35_{022}8_{1\frac{1}{2}\frac{1}{2}},\\\nonumber
&64_{1\frac{3}{2}\frac{3}{2}}&=-\frac{2\sqrt{7}}{35}35_{222}8_{-1\frac{1}{2}-\frac{1}{2}}
+\frac{\sqrt{7}}{35}35_{221}8_{-1\frac{1}{2}\frac{1}{2}}+\frac{2\sqrt{7}}{35}35_{1\frac{5}{2}\frac{5}{2}}8_{01-1}
-\frac{2\sqrt{70}}{175}35_{1\frac{5}{2}\frac{3}{2}}8_{010}+\frac{\sqrt{70}}{175}35_{1\frac{5}{2}\frac{1}{2}}8_{011}
\\\nonumber
&+&\frac{\sqrt{105}}{25}35_{1\frac{3}{2}\frac{3}{2}}8_{010}
-\frac{3\sqrt{35}}{35}35_{1\frac{3}{2}\frac{3}{2}}8_{000}
-\frac{\sqrt{70}}{25}35_{1\frac{3}{2}\frac{1}{2}}8_{011}+\frac{6\sqrt{7}}{35}35_{022}8_{1\frac{1}{2}-\frac{1}{2}}
-\frac{3\sqrt{7}}{35}35_{021}8_{1\frac{1}{2}\frac{1}{2}}+\frac{\sqrt{7}}{7}35_{011}8_{1\frac{1}{2}\frac{1}{2}},\\\nonumber
&28_{233}&=\frac{\sqrt{2}}{2}35_{222}8_{011}-\frac{\sqrt{2}}{2}35_{1\frac{5}{2}\frac{5}{2}}8_{1\frac{1}{2}\frac{1}{2}},\\\nonumber
&28_{1\frac{5}{2}\frac{5}{2}}&=\frac{\sqrt{3}}{6}35_{222}8_{-1\frac{1}{2}\frac{1}{2}}
-\frac{\sqrt{6}}{12}35_{1\frac{5}{2}\frac{5}{2}}8_{010}
-\frac{\sqrt{2}}{4}35_{1\frac{5}{2}\frac{5}{2}}8_{000}
+\frac{\sqrt{15}}{30}35_{1\frac{5}{2}\frac{3}{2}}8_{011}+\frac{\sqrt{10}}{5}35_{1\frac{3}{2}\frac{3}{2}}8_{011}
-\frac{\sqrt{3}}{3}35_{022}8_{1\frac{1}{2}\frac{1}{2}},\\\nonumber
&35^{(1)}_{222}&=\frac{2}{3}35_{222}8_{010}+\frac{2\sqrt{3}}{9}35_{222}8_{000}
-\frac{\sqrt{2}}{3}35_{221}8_{011}
-\frac{\sqrt{15}}{9}35_{1\frac{3}{2}\frac{3}{2}}8_{1\frac{1}{2}\frac{1}{2}},\\\nonumber
&35^{(1)}_{1\frac{5}{2}\frac{5}{2}}&=\frac{2}{3}35_{1\frac{5}{2}\frac{5}{2}}8_{010}
+\frac{2\sqrt{3}}{9}35_{1\frac{5}{2}\frac{5}{2}}8_{000}
-\frac{2\sqrt{10}}{15}35_{1\frac{5}{2}\frac{3}{2}}8_{011}+\frac{\sqrt{15}}{45}35_{1\frac{3}{2}\frac{3}{2}}8_{011}
-\frac{\sqrt{2}}{3}35_{022}8_{1\frac{1}{2}\frac{1}{2}},\\\nonumber
&35^{(1)}_{1\frac{3}{2}\frac{3}{2}}&=\frac{\sqrt{15}}{9}35_{222}8_{-1\frac{1}{2}-\frac{1}{2}}
-\frac{\sqrt{15}}{18}35_{221}8_{-1\frac{1}{2}\frac{1}{2}}-\frac{\sqrt{15}}{45}35_{1\frac{5}{2}\frac{5}{2}}8_{01-1}
+\frac{\sqrt{6}}{45}35_{1\frac{5}{2}\frac{3}{2}}8_{010}-\frac{\sqrt{6}}{90}35_{1\frac{5}{2}\frac{1}{2}}8_{011}
+\frac{31}{60}35_{1\frac{3}{2}\frac{3}{2}}8_{010}\\\nonumber
&+&\frac{\sqrt{3}}{12}35_{1\frac{3}{2}\frac{3}{2}}8_{000}
-\frac{31\sqrt{6}}{180}35_{1\frac{3}{2}\frac{1}{2}}8_{011}+\frac{\sqrt{15}}{90}35_{022}8_{1\frac{1}{2}-\frac{1}{2}}
-\frac{\sqrt{15}}{180}35_{021}8_{1\frac{1}{2}\frac{1}{2}}-\frac{5\sqrt{15}}{36}35_{011}8_{1\frac{1}{2}\frac{1}{2}},\\\nonumber
&35^{(1)}_{022}&=\frac{\sqrt{2}}{3}35_{1\frac{5}{2}\frac{5}{2}}8_{-1\frac{1}{2}-\frac{1}{2}}
-\frac{\sqrt{10}}{15}35_{1\frac{5}{2}\frac{3}{2}}8_{-1\frac{1}{2}\frac{1}{2}}
+\frac{\sqrt{15}}{90}35_{1\frac{3}{2}\frac{3}{2}}8_{-1\frac{1}{2}\frac{1}{2}}
+\frac{1}{2}35_{022}8_{010}+\frac{\sqrt{3}}{18}35_{022}8_{000}-\frac{\sqrt{2}}{4}35_{021}8_{011}\\\nonumber
&+&\frac{\sqrt{2}}{12}35_{011}8_{011}
-\frac{\sqrt{3}}{3}35_{-1\frac{3}{2}\frac{3}{2}}8_{1\frac{1}{2}\frac{1}{2}},
\end{eqnarray}
\begin{eqnarray}\nonumber
&35^{(1)}_{011}&=\frac{5\sqrt{15}}{36}35_{1\frac{3}{2}\frac{3}{2}}8_{-1\frac{1}{2}-\frac{1}{2}}-\frac{5\sqrt{15}}{36}35_{1\frac{3}{2}\frac{1}{2}}8_{-1\frac{1}{2}\frac{1}{2}}
-\frac{\sqrt{2}}{12}35_{022}8_{01-1} +\frac{1}{12}35_{021}8_{010}
-\frac{\sqrt{3}}{36}35_{020}8_{011}
+\frac{13}{36}35_{011}8_{010}\\\nonumber
&-&\frac{\sqrt{3}}{18}35_{011}8_{000}
-\frac{13}{36}35_{010}8_{011}
+\frac{\sqrt{3}}{18}35_{-1\frac{3}{2}\frac{3}{2}}8_{1\frac{1}{2}-\frac{1}{2}}
-\frac{1}{18}35_{-1\frac{3}{2}\frac{1}{2}}8_{1\frac{1}{2}\frac{1}{2}}-\frac{5}{9}35_{-1\frac{1}{2}\frac{1}{2}}8_{1\frac{1}{2}\frac{1}{2}},\\\nonumber
&35^{(1)}_{-1\frac{3}{2}\frac{3}{2}}&=\frac{\sqrt{3}}{3}35_{022}8_{-1\frac{1}{2}-\frac{1}{2}}
-\frac{\sqrt{3}}{6}35_{021}8_{-1\frac{1}{2}\frac{1}{2}}+\frac{\sqrt{3}}{18}35_{011}8_{-1\frac{1}{2}\frac{1}{2}}
+\frac{1}{3}35_{-1\frac{3}{2}\frac{3}{2}}8_{010}-\frac{\sqrt{3}}{9}35_{-1\frac{3}{2}\frac{3}{2}}8_{000}
-\frac{\sqrt{6}}{9}35_{-1\frac{3}{2}\frac{1}{2}}8_{011}\\\nonumber
&+&\frac{\sqrt{6}}{18}35_{-1\frac{1}{2}\frac{1}{2}}8_{011}-\frac{\sqrt{3}}{3}35_{-211}8_{1\frac{1}{2}\frac{1}{2}},\\\nonumber
&35^{(1)}_{-1\frac{1}{2}\frac{1}{2}}&=\frac{5}{9}35_{011}8_{-1\frac{1}{2}-\frac{1}{2}}
-\frac{5\sqrt{2}}{18}35_{010}8_{-1\frac{1}{2}\frac{1}{2}}
-\frac{\sqrt{6}}{18}35_{-1\frac{3}{2}\frac{3}{2}}8_{01-1}
+\frac{1}{9}35_{-1\frac{3}{2}\frac{1}{2}}8_{010}-\frac{\sqrt{2}}{18}35_{-1\frac{3}{2}-\frac{1}{2}}8_{011}\\\nonumber
&+&\frac{7}{36}35_{-1\frac{1}{2}\frac{1}{2}}8_{010}-\frac{7\sqrt{3}}{36}35_{-1\frac{1}{2}\frac{1}{2}}8_{000}
-\frac{7\sqrt{2}}{36}35_{-1\frac{1}{2}-\frac{1}{2}}8_{011}+\frac{1}{6}35_{-211}8_{1\frac{1}{2}-\frac{1}{2}}
-\frac{\sqrt{2}}{12}35_{-210}8_{1\frac{1}{2}\frac{1}{2}}-\frac{5\sqrt{3}}{18}35_{-200}8_{1\frac{1}{2}\frac{1}{2}},\\\nonumber
&35^{(1)}_{-211}&=\frac{\sqrt{3}}{3}35_{-1\frac{3}{2}\frac{3}{2}}8_{-1\frac{1}{2}-\frac{1}{2}}
-\frac{1}{3}35_{-1\frac{3}{2}\frac{1}{2}}8_{-1\frac{1}{2}\frac{1}{2}}
+\frac{1}{6}35_{-1\frac{1}{2}\frac{1}{2}}8_{-1\frac{1}{2}\frac{1}{2}}
+\frac{1}{6}35_{-211}8_{010}-\frac{5\sqrt{3}}{18}35_{-211}8_{000}-\frac{1}{6}35_{-210}8_{011}\\\nonumber
&+&\frac{\sqrt{6}}{18}35_{-200}8_{011}
-\frac{\sqrt{2}}{3}35_{-3\frac{1}{2}\frac{1}{2}}8_{1\frac{1}{2}\frac{1}{2}},\\\nonumber
&35^{(1)}_{-200}&=\frac{5\sqrt{3}}{18}35_{-1\frac{1}{2}\frac{1}{2}}8_{-1\frac{1}{2}-\frac{1}{2}}
-\frac{5\sqrt{3}}{18}35_{-1\frac{1}{2}-\frac{1}{2}}8_{-1\frac{1}{2}\frac{1}{2}}
-\frac{\sqrt{6}}{18}35_{-211}8_{01-1}+\frac{\sqrt{6}}{18}35_{-210}8_{010}-\frac{\sqrt{6}}{18}35_{-21-1}8_{011}\\\nonumber
&-&\frac{\sqrt{3}}{3}35_{-200}8_{000}
+\frac{\sqrt{6}}{9}35_{-3\frac{1}{2}\frac{1}{2}}8_{1\frac{1}{2}-\frac{1}{2}}-\frac{\sqrt{6}}{9}35_{-3\frac{1}{2}-\frac{1}{2}}8_{1\frac{1}{2}\frac{1}{2}},\\\nonumber
&35^{(1)}_{-3\frac{1}{2}\frac{1}{2}}&=\frac{\sqrt{2}}{3}35_{-211}8_{-1\frac{1}{2}-\frac{1}{2}}
-\frac{1}{3}35_{-210}8_{-1\frac{1}{2}\frac{1}{2}}+\frac{\sqrt{6}}{9}35_{-200}8_{-1\frac{1}{2}\frac{1}{2}}
-\frac{4\sqrt{3}}{9}35_{-3\frac{1}{2}\frac{1}{2}}8_{000},\\\nonumber
&35^{(2)}_{222}&=-\frac{\sqrt{2}}{6}35_{222}8_{010}+\frac{7\sqrt{6}}{36}35_{222}8_{000}
+\frac{1}{6}35_{221}8_{011}+\frac{3}{4}35_{1\frac{5}{2}\frac{5}{2}}8_{1\frac{1}{2}-\frac{1}{2}}
-\frac{3\sqrt{5}}{20}35_{1\frac{5}{2}\frac{3}{2}}8_{1\frac{1}{2}\frac{1}{2}}
-\frac{\sqrt{30}}{45}35_{1\frac{3}{2}\frac{3}{2}}8_{1\frac{1}{2}\frac{1}{2}},\\\nonumber
&35^{(2)}_{1\frac{5}{2}\frac{5}{2}}&=\frac{3}{4}35_{222}8_{-1\frac{1}{2}\frac{1}{2}}
+\frac{5\sqrt{2}}{24}35_{1\frac{5}{2}\frac{5}{2}}8_{010}
-\frac{13\sqrt{6}}{72}35_{1\frac{5}{2}\frac{5}{2}}8_{000}
-\frac{\sqrt{5}}{12}35_{1\frac{5}{2}\frac{3}{2}}8_{011}-\frac{\sqrt{30}}{18}35_{1\frac{3}{2}\frac{3}{2}}8_{011}
+\frac{1}{6}35_{022}8_{1\frac{1}{2}\frac{1}{2}},\\\nonumber
&35^{(2)}_{1\frac{3}{2}\frac{3}{2}}&=\frac{\sqrt{30}}{45}35_{222}8_{-1\frac{1}{2}-\frac{1}{2}}
-\frac{\sqrt{30}}{90}35_{221}8_{-1\frac{1}{2}\frac{1}{2}}+\frac{\sqrt{30}}{18}35_{1\frac{5}{2}\frac{5}{2}}8_{01-1}
-\frac{\sqrt{3}}{9}35_{1\frac{5}{2}\frac{3}{2}}8_{010}+\frac{\sqrt{3}}{18}35_{1\frac{5}{2}\frac{1}{2}}8_{011}
-\frac{\sqrt{2}}{6}35_{1\frac{3}{2}\frac{3}{2}}8_{010}\\\nonumber
&+&\frac{\sqrt{6}}{6}35_{1\frac{3}{2}\frac{3}{2}}8_{000}
+\frac{\sqrt{3}}{9}35_{1\frac{3}{2}\frac{1}{2}}8_{011}+\frac{11\sqrt{30}}{90}35_{022}8_{1\frac{1}{2}-\frac{1}{2}}
-\frac{11\sqrt{30}}{180}35_{021}8_{1\frac{1}{2}\frac{1}{2}}-\frac{\sqrt{30}}{36}35_{011}8_{1\frac{1}{2}\frac{1}{2}},\\\nonumber
&35^{(2)}_{022}&=-\frac{1}{6}35_{1\frac{5}{2}\frac{5}{2}}8_{-1\frac{1}{2}-\frac{1}{2}}
+\frac{\sqrt{5}}{30}35_{1\frac{5}{2}\frac{3}{2}}8_{-1\frac{1}{2}\frac{1}{2}}
+\frac{11\sqrt{30}}{90}35_{1\frac{3}{2}\frac{3}{2}}8_{-1\frac{1}{2}\frac{1}{2}}
+\frac{\sqrt{2}}{4}35_{022}8_{010}-\frac{5\sqrt{6}}{36}35_{022}8_{000}-\frac{1}{4}35_{021}8_{011}\\\nonumber
&-&\frac{5}{12}35_{011}8_{011}
+\frac{\sqrt{6}}{12}35_{-1\frac{3}{2}\frac{3}{2}}8_{1\frac{1}{2}\frac{1}{2}},\\\nonumber
&35^{(2)}_{011}&=\frac{\sqrt{30}}{36}35_{1\frac{3}{2}\frac{3}{2}}8_{-1\frac{1}{2}-\frac{1}{2}}
-\frac{\sqrt{10}}{36}35_{1\frac{3}{2}\frac{1}{2}}8_{-1\frac{1}{2}\frac{1}{2}}
+\frac{5}{12}35_{022}8_{01-1} -\frac{5\sqrt{2}}{24}35_{021}8_{010}
+\frac{5\sqrt{6}}{72}35_{020}8_{011}
-\frac{11\sqrt{2}}{72}35_{011}8_{010}\\\nonumber
&+&\frac{5\sqrt{6}}{36}35_{011}8_{000}
+\frac{11\sqrt{2}}{72}35_{010}8_{011}
+\frac{17\sqrt{6}}{72}35_{-1\frac{3}{2}\frac{3}{2}}8_{1\frac{1}{2}-\frac{1}{2}}
-\frac{17\sqrt{2}}{72}35_{-1\frac{3}{2}\frac{1}{2}}8_{1\frac{1}{2}\frac{1}{2}}
-\frac{\sqrt{2}}{9}35_{-1\frac{1}{2}\frac{1}{2}}8_{1\frac{1}{2}\frac{1}{2}},\\\nonumber
&35^{(2)}_{-1\frac{3}{2}\frac{3}{2}}&=-\frac{\sqrt{6}}{12}35_{022}8_{-1\frac{1}{2}-\frac{1}{2}}
+\frac{\sqrt{6}}{24}35_{021}8_{-1\frac{1}{2}\frac{1}{2}}
+\frac{17\sqrt{6}}{72}35_{011}8_{-1\frac{1}{2}\frac{1}{2}}
+\frac{7\sqrt{2}}{24}35_{-1\frac{3}{2}\frac{3}{2}}8_{010}
-\frac{7\sqrt{6}}{72}35_{-1\frac{3}{2}\frac{3}{2}}8_{000}\\\nonumber
&-&\frac{7\sqrt{3}}{36}35_{-1\frac{3}{2}\frac{1}{2}}8_{011}
-\frac{5\sqrt{3}}{18}35_{-1\frac{1}{2}\frac{1}{2}}8_{011}
+\frac{\sqrt{6}}{12}35_{-211}8_{1\frac{1}{2}\frac{1}{2}},\\\nonumber
&35^{(2)}_{-1\frac{1}{2}\frac{1}{2}}&=\frac{\sqrt{2}}{9}35_{011}8_{-1\frac{1}{2}-\frac{1}{2}}
-\frac{1}{9}35_{010}8_{-1\frac{1}{2}\frac{1}{2}}
+\frac{5\sqrt{3}}{18}35_{-1\frac{3}{2}\frac{3}{2}}8_{01-1}
-\frac{5\sqrt{2}}{18}35_{-1\frac{3}{2}\frac{1}{2}}8_{010}
+\frac{5}{18}35_{-1\frac{3}{2}-\frac{1}{2}}8_{011}\\\nonumber
&-&\frac{\sqrt{2}}{9}35_{-1\frac{1}{2}\frac{1}{2}}8_{010}
+\frac{\sqrt{6}}{9}35_{-1\frac{1}{2}\frac{1}{2}}8_{000}
+\frac{2}{9}35_{-1\frac{1}{2}-\frac{1}{2}}8_{011}
+\frac{\sqrt{2}}{3}35_{-211}8_{1\frac{1}{2}-\frac{1}{2}}
-\frac{1}{3}35_{-210}8_{1\frac{1}{2}\frac{1}{2}}
-\frac{\sqrt{6}}{18}35_{-200}8_{1\frac{1}{2}\frac{1}{2}},\\\nonumber
&35^{(2)}_{-211}&=-\frac{\sqrt{6}}{12}35_{-1\frac{3}{2}\frac{3}{2}}8_{-1\frac{1}{2}-\frac{1}{2}}
+\frac{\sqrt{2}}{12}35_{-1\frac{3}{2}\frac{1}{2}}8_{-1\frac{1}{2}\frac{1}{2}}
+\frac{\sqrt{2}}{3}35_{-1\frac{1}{2}\frac{1}{2}}8_{-1\frac{1}{2}\frac{1}{2}}
+\frac{\sqrt{2}}{3}35_{-211}8_{010}
-\frac{\sqrt{6}}{18}35_{-211}8_{000}\\\nonumber
&-&\frac{\sqrt{2}}{3}35_{-210}8_{011}
-\frac{5\sqrt{3}}{18}35_{-200}8_{011}
+\frac{1}{6}35_{-3\frac{1}{2}\frac{1}{2}}8_{1\frac{1}{2}\frac{1}{2}},\\\nonumber
&35^{(2)}_{-200}&=\frac{\sqrt{6}}{18}35_{-1\frac{1}{2}\frac{1}{2}}8_{-1\frac{1}{2}-\frac{1}{2}}
-\frac{\sqrt{6}}{18}35_{-1\frac{1}{2}-\frac{1}{2}}8_{-1\frac{1}{2}\frac{1}{2}}
+\frac{5\sqrt{3}}{18}35_{-211}8_{01-1}
-\frac{5\sqrt{3}}{18}35_{-210}8_{010}
+\frac{5\sqrt{3}}{18}35_{-21-1}8_{011}\\\nonumber
&+&\frac{\sqrt{6}}{12}35_{-200}8_{000}
+\frac{7\sqrt{3}}{36}35_{-3\frac{1}{2}\frac{1}{2}}8_{1\frac{1}{2}-\frac{1}{2}}
-\frac{7\sqrt{3}}{36}35_{-3\frac{1}{2}-\frac{1}{2}}8_{1\frac{1}{2}\frac{1}{2}},\\\nonumber
&35^{(2)}_{-3\frac{1}{2}\frac{1}{2}}&=-\frac{1}{6}35_{-211}8_{-1\frac{1}{2}-\frac{1}{2}}
+\frac{\sqrt{2}}{12}35_{-210}8_{-1\frac{1}{2}\frac{1}{2}}
+\frac{7\sqrt{3}}{36}35_{-200}8_{-1\frac{1}{2}\frac{1}{2}}
+\frac{3\sqrt{2}}{8}35_{-3\frac{1}{2}\frac{1}{2}}8_{010}
-\frac{\sqrt{6}}{72}35_{-3\frac{1}{2}\frac{1}{2}}8_{000}\\\nonumber
&-&\frac{3}{4}35_{-3\frac{1}{2}-\frac{1}{2}}8_{011}.
\end{eqnarray}
And others can be obtain by the action of $I_{\pm}$ and $K_{\pm}$
on given decomposition and the action rules on
$\mu^{(\lambda)}_{(\nu)}$ are as follows
\begin{eqnarray}
\mathcal{O}\mu^{(\lambda)}_{(\nu)}=\sum\limits_{\nu_1,\nu_2}
\left(\begin{array}{ccc}\mu_1&\mu_2&\mu\\\nu_1&\nu_2&\nu\end{array}\right)\left[\left(\mathcal{O}\mu_{1(\nu_1)}\right)\mu_{2(\nu_2)}
+\mu_{1(\nu_1)}\left(\mathcal{O}\mu_{2(\nu_2)}\right)\right],
\end{eqnarray}
where $\mathcal{O}$ denotes any one operator in
(\ref{Ipm})-(\ref{Lpm}).
\end{widetext}

\end{document}